\documentclass[prl,twocolumn,showpacs]{revtex4-1}
\usepackage{graphicx}

\begin{document}
\title{Inseparability criteria based on bipartitions of $N$-spin systems}
\author{Asoka Biswas}
\affiliation{Department of Physcs, Indian Institute of Technology Ropar, Rupnagar, Punjab 140001, India}
\date{\today}
\begin{abstract}
We present a new set of inseparabilty inequalities to detect entanglement in $N$-spin states. These are based on negative partial transposition and involve collective spin-spin correlations of any two partitions of the entire system. They reveal the rich texture of partial separability for different partitions and can discriminate between GHZ-type and W-type entanglement, as well. We introduce a new concept of relative entanglement of two different systems and two different partitions in a spin-ensemble. These criteria can be equally applicable to non-symmetric states and states with odd or even $N$.
\end{abstract}
\maketitle

In recent times, entanglement in many-spin systems has attracted substantial attention in view of scalable quantum computation, which relies on such entanglement as a resource. Therefore, it is important to identify at the outset of an experiment whether the state under consideration is entangled or not.
However, nature of entanglement in large ensembles is not yet well understood. It is customary to look into entanglement between different partitions in such case. States of a few spins, for example, are classified in terms of separability with respect to different bipartitions. It is shown in \cite{cirac99}, that a state of 3 spins A, B, and C is 1-spin biseparable, if it can be written as, viz.,  $\rho=\sum_i\rho_i^A\otimes\rho_i^{BC}$ and cannot be written in similar forms with respect to the other spins B and C.  Similarly, for $N(>3)$-spin system, one can have $m$-spin biseparability ($m=2,\ldots,N$), where such biseparability can be obtained for at most $m$ number of spins.  To identify entanglement in such bipartitions, the Peres-Horodecki criterion has been most successful \cite{peres,horo}. According to this criterion, a separable density matrix $\rho$ under partial transposition remains non-negative.  While this is a necessary and sufficient condition for inseparability in 2-qubit and qubit-qutrit systems, it poses experimental challenge for large ensembles of $N$ spins ($N\gg 1$), as it demands quantum state tomographic measurement of all the elements of $\rho$.  Moreover, this criterion cannot distinguish two inequivalent classes of entangled states of $N\ge 3$ spins, namely, GHZ-type and W-type, which cannot be transformed into each other using local operation and classical communication.  
We note that there exist other inseparability criteria, e.g., based on covariance matrix, that reveal whether two partitions of a large ensemble are entangled or not \cite{guhne}. These test the negativity of the off-diagonal block $C$ of the covariance matrix. However, to verify this in experiments, one needs to measure all the elements of the matrix $C$, and it often involves single-spin addressing \cite{raja}. So scalability to large ensembles, where single spin addressing is not  feasible, is not straightforward. They also cannot differentiate GHZ- or W-type entanglement. In addition, these criteria are suitable only for symmetric states and the states with even numbers of spins \cite{raja1}. In this paper, we introduce a new set of criteria that overcome the issues of experimental challenges.  They do not require single spin addressing and involve higher order moments of collective spin components of different bipartitions \cite{biswas}. They can distinguish between different inequivalent classes of multipartite entanglement and also reveal entanglement in {\it all quantum states irrespective of its symmetry or number of spins\/}. The advantage of using collective spin operators instead of single-spin operators is that this unifies and therefore simplifies the inseparability criteria to a great extent. These criteria also lead us to a new insight of partition-dependent entanglement, that describes how a state becomes differently entangled amongst different partitions. To this end, we put forward a new measure of entanglement, {\it for the first time\/}, based on how strongly a quantum state satisfies these criteria.

Another important inseparability criteria, based on spin squeezing \cite{toth}, are given as a function of  collective spin components.  However they detect genuine entanglement in the state in terms of inseparability of the averaged $m$-spin reduced density matrix. In contrast, our criteria can determine the existence of partial separability  and can also distinguish between different classes of genuine entanglement.
Moreover, they  are quite distinct from those, based on, e.g., Schrodinger-Robertson uncertainty relation \cite{gsa}, entropic uncertainty relations \cite{guhne1}, positive partial transposition of multi-spin states \cite{raja2}, and realignment of density matrix \cite{rudolph}. 


In the following, we derive our criteria based on Shchukin-Vogel (SV) criterion of inseparability \cite{sv} that was originally proposed for two-mode continuous-variable entanglement. To start with, let us consider a state $\rho$ of a $N$-spin system. The state $\rho$ is positive semidefinite and satisfies Tr $(\rho \hat{f}^\dag \hat{f}) \ge 0$, for any operator $\hat{f}$, whose normally ordered form exists \cite{sv}. We now consider two partitions A and B of the entire system, with respective number of spins $n_A$ and $n_B$, such that $N=n_A+n_B$. The above property of $\rho$ suggests that that the partial transpose of $\rho$ with respect to one of the partitions (say, B) will also be positive semidefinite, if
 \begin{equation}
{\rm Tr}(\rho^{\rm T_B}\hat{f}^\dag\hat{f}) \ge 0\label{ineq1} \;.
\label{c1}
\end{equation}
This condition holds, iff $\rho^{\rm T_B}$ is a legitimate density matrix. According to Peres criterion, therefore, $\rho$ is a state that is biseparable into two partitions A and B. We next choose a normally ordered form of $\hat{f}$, as
\begin{equation}
\hat{f}=\sum_{klmn}C_{klmn}S_+^{A^k}S_-^{A^l}S_+^{B^m}S_-^{B^n}\;,
\label{f}
\end{equation}
where $S_\pm$ represent the collective spin creation and annihilation operators of the respective partitions, e.g.,
\begin{equation}S_\pm^A=\sum_{i=1}^{n_A}s_\pm^i\;, ~s_\pm^i\equiv s_\pm^i\otimes_{j=1(\ne i)}^{n_A}{\mathbf 1}_j\;.
\end{equation}
Note that any anti-normal arrangement of $S_\pm$ can be transformed into normal ordering using $[S_+,S_-]=2m$ as
\begin{equation}
S_-^jS_+^k=\sum_{r=0}^{{\rm min}(j,k)}\frac{j!k!(-2m)^r}{r!(j-r)!(k-r)!}S_+^{j-r}S_-^{k-r}\;,
\label{antinormal}
\end{equation}
when it applies to the state $|S,m\rangle$, $s$ being the total spin quantum number of the partition and $m$ being its projection along quantization axis.
The operator $\hat{f}^\dag \hat{f}$ can be written as
\begin{eqnarray}
\hat{f}^\dag \hat{f}&=&\sum_{klmn,pqrs}\left[C_{pqrs}^*C_{klmn}\right.\nonumber\\
&&\left.\left(S_+^{A^q}S_-^{A^p}S_+^{A^k}S_-^{A^l}\right)\left(S_+^{B^s}S_-^{B^r}S_+^{B^m}S_-^{B^n}\right)\right]\;.
\end{eqnarray}
Because the partial transposition is a local positive operation, it maintains the symmetry for two operators $X$ and $Y$ as Tr$(X^{\rm PT}Y)$=Tr$(XY^{\rm PT})$. Therefore, our criterion (\ref{c1}) for separability can be written as
\begin{equation}
{\rm Tr}\left[\left(\hat{f}^\dag \hat{f}\right)^{\rm T_B}\rho\right]\ge 0\;,
\label{ineq2}
\end{equation}
\begin{eqnarray}
\left(\hat{f}^\dag \hat{f}\right)^{\rm T_B}&=&\sum_{klmn,pqrs}\left[C_{pqrs}^*C_{klmn}\right.\nonumber\\
&&\left.\left(S_+^{A^q}S_-^{A^p}S_+^{A^k}S_-^{A^l}\right)\left(S_+^{B^n}S_-^{B^m}S_+^{B^r}S_-^{B^s}\right)\right]\;.
\label{pt}
\end{eqnarray}
This criterion should hold for arbitrary vector of coefficients $C_{klmn}$. So, Eq. (\ref{ineq2}) is equivalent to positive semi-definiteness of the matrix with the following moments as elements:
\begin{equation}
M_{klmn,pqrs}=\left\langle \left(S_+^{A^q}S_-^{A^p}S_+^{A^k}S_-^{A^l}\right)\left(S_+^{B^n}S_-^{B^m}S_+^{B^r}S_-^{B^s}\right)\right\rangle_\rho\;.
\label{Matrix}
\end{equation}
According to Sylvester's criterion \cite{sylvester}, the matrix $M$ is positive semi-definite iff all its principal minors are non-negative. We can thus have the following:

{\it Inseparability criterion: A state $\rho$ is inseparable for the partitions A and B, if the matrix $M$, with elements given by Eq. (\ref{Matrix}) is negative, or in other words, if there exists any principal minor of the matrix $M$, that is negative.}

With a systematic choice of the indices $(k,l,m,n; p,q,r,s)$ we obtain the following form of the matrix $M$:
\begin{widetext}
\begin{equation}
M = \left[ {\begin{array}{*{20}{c}}
1&{\left\langle {S_ + ^B} \right\rangle }&{\left\langle {S_ - ^B} \right\rangle }&{\left\langle {S_ - ^A} \right\rangle }&{\left\langle {S_ + ^A} \right\rangle }&{\left\langle {S_ + ^{{B^2}}} \right\rangle }&{\left\langle {S_ + ^BS_ - ^B} \right\rangle }&{\left\langle {S_ - ^AS_ + ^B} \right\rangle }&{\left\langle {S_ + ^AS_ + ^B} \right\rangle }& \cdots \\
{\left\langle {S_ - ^B} \right\rangle }&{\left\langle {S_ + ^BS_ - ^B} \right\rangle }&{\left\langle {S_ - ^{{B^2}}} \right\rangle }&{\left\langle {S_ - ^AS_ - ^B} \right\rangle }&{\left\langle {S_ + ^AS_ - ^B} \right\rangle }&{\left\langle {S_ + ^{{B^2}}S_ - ^B} \right\rangle }&{\left\langle {S_ + ^BS_ - ^{{B^2}}} \right\rangle }&{\left\langle {S_ - ^AS_ + ^BS_ - ^B} \right\rangle }&{\left\langle {S_ + ^AS_ + ^BS_ - ^B} \right\rangle }& \cdots \\
{\left\langle {S_ + ^B} \right\rangle }&{\left\langle {S_ + ^{{B^2}}} \right\rangle }&{\left\langle {S_ - ^BS_ + ^B} \right\rangle }&{\left\langle {S_ - ^AS_ + ^B} \right\rangle }&{\left\langle {S_ + ^AS_ + ^B} \right\rangle }&{\left\langle {S_ + ^{{B^3}}} \right\rangle }&{\left\langle {S_ + ^BS_ - ^BS_ + ^B} \right\rangle }&{\left\langle {S_ - ^AS_ + ^{{B^2}}} \right\rangle }&{\left\langle {S_ + ^AS_ + ^{{B^2}}} \right\rangle }& \cdots \\
{\left\langle {S_ + ^A} \right\rangle }&{\left\langle {S_ + ^AS_ + ^B} \right\rangle }&{\left\langle {S_ + ^AS_ - ^B} \right\rangle }&{\left\langle {S_ + ^AS_ - ^A} \right\rangle }&{\left\langle {S_ + ^{{A^2}}} \right\rangle }&{\left\langle {S_ + ^AS_ + ^{{B^2}}} \right\rangle }&{\left\langle {S_ + ^AS_ + ^BS_ - ^B} \right\rangle }&{\left\langle {S_ + ^AS_ - ^AS_ + ^B} \right\rangle }&{\left\langle {S_ + ^{{A^2}}S_ + ^B} \right\rangle }& \cdots \\
{\left\langle {S_ - ^A} \right\rangle }&{\left\langle {S_ - ^AS_ + ^B} \right\rangle }&{\left\langle {S_ - ^AS_ - ^B} \right\rangle }&{\left\langle {S_ - ^{{A^2}}} \right\rangle }&{\left\langle {S_ - ^AS_ + ^A} \right\rangle }&{\left\langle {S_ - ^AS_ + ^{{B^2}}} \right\rangle }&{\left\langle {S_ - ^AS_ + ^BS_ - ^B} \right\rangle }&{\left\langle {S_ - ^{{A^2}}S_ + ^B} \right\rangle }&{\left\langle {S_ - ^AS_ + ^AS_ + ^B} \right\rangle }& \cdots \\
 \vdots & \vdots & \vdots & \vdots & \vdots & \vdots & \vdots & \vdots & \vdots & \vdots
\end{array}} \right]
\end{equation}
\end{widetext}
Note that the principal minor corresponds to the determinant of a $k\times k$ submatrix $M'$ of $M$. $M'$ has the following elements: $M'_{i,j}=M_{\alpha_i,\alpha_j}, i,j=1,2,\ldots,k$, arranged in such a way that $\alpha_i\le\alpha_j$ if $i\le j$.

We find that there exist two sets of such submatrices, that are negative for two inequivalent classes of pure entangled states, namely {\it Class I}: GHZ-type and {\it Class II}: W-type, and are given by:
\begin{widetext}
\begin{eqnarray}
{\rm Class ~I}: & -P_I(N,n_A)=\left| {\begin{array}{*{20}{c}}
{\left\langle {S_ + ^{{A^{(n_A - 1)}}}S_ - ^{{A^{(n_A - 1)}}}S_ - ^{{B^{(n_B - 1)}}}S_ + ^{{B^{(n_B - 1)}}}} \right\rangle }&{\left\langle {S_ + ^{{A^{n_A}}}S_ + ^{{B^{n_B}}}} \right\rangle }\\
{\left\langle {S_ - ^{{A^{n_A}}}S_ - ^{{B^{n_B}}}} \right\rangle }&{\left\langle {S_ - ^AS_ + ^AS_ + ^BS_ - ^B} \right\rangle }
\end{array}} \right| < 0\;; \label{class1}\\
{\rm Class~ II}: & -P_{II}(N,n_A)=\left| {\begin{array}{*{20}{c}}
1&{\left\langle {S_ - ^AS_ + ^B} \right\rangle }\\
{\left\langle {S_ + ^AS_ - ^B} \right\rangle }&{\left\langle {S_ + ^AS_ - ^AS_ + ^BS_ - ^B} \right\rangle }
\end{array}} \right|  < 0\;.\label{class2}
\end{eqnarray}
\end{widetext}
Our proposition now is as follows: {\it If the above inequalities are satisfied for all possible bi-partitions, the state under consideration is fully inseparable in one of the classes, otherwise partially separable.} Clearly, this constitutes a {\it sufficient} criterion for inseparability. We further find that for a symmetric state, it is enough to consider any of the partitions to detect genuine entanglement. Below, we illustrate our criterion for several states:

{\it Example 1:} A pure 3-qubit GHZ state $|\psi\rangle=(|000\rangle+|111\rangle)/\sqrt{2}$: Considering first two spins in partition A (i.e., $n_A=2$), we can rewrite the state in collective spin basis of each partition as $|\psi\rangle=(|1,-1\rangle_A |1/2,-1/2\rangle_B+|1,+1\rangle_A |1/2,+1/2\rangle_B)/\sqrt{2}$. The Class I inequality (\ref{class1}) leads to
\begin{eqnarray}
P_I(3,2)&=&\left\langle {S_ + ^{{A^2}}S_ + ^B} \right\rangle \left\langle {S_ - ^{{A^2}}S_ - ^B} \right\rangle-\left\langle {S_ + ^AS_ - ^A} \right\rangle \left\langle {S_ - ^AS_ + ^AS_ + ^BS_ - ^B} \right\rangle   \nonumber\\
&=&  1 > 0\;,
\end{eqnarray}
while the other inequaility (\ref{class2}) gives $P_{II}(3,2)<0$. Clearly, this signals existence of GHZ-type inseparability and not W-type inseparability. It is not required to consider any other partition to test the same, as the state is symmetric.

{\it Example 2:} A pure 3-qubit W-state $|\psi\rangle=(|001\rangle+|010\rangle+|100\rangle)/\sqrt{3}$: Considering again first two spins in partition A and rewriting the state in collective spin basis as $\left| \psi  \right\rangle  = \left[ {{{\left| {1, - 1} \right\rangle }_A}{{\left| 1 \right\rangle }_B} + \sqrt 2 {{\left| {1,0} \right\rangle }_A}{{\left| 0 \right\rangle }_B}} \right]/\sqrt{3}$, we obtain from Class II inequality (\ref{class2})
\begin{eqnarray}
P_{II}(3,2)&=&\left\langle {S_ + ^AS_ - ^B} \right\rangle \left\langle {S_ - ^AS_ + ^B} \right\rangle-\left\langle {S_ + ^AS_ - ^AS_ + ^BS_ - ^B} \right\rangle   \nonumber\\
& =&  \frac{4}{9} > 0\;,
\label{class2a}
\end{eqnarray}
while Eq. (\ref{class1}) gives $P_I(3,2)<0$, signallng W-type inseparability \cite{hillery_zubairy} and not GHZ-type.

The above result can be generalized to pure $N$-spin inseparable states. For a GHZ-type state $\left| \psi  \right\rangle  = \cos \theta {\left| 0 \right\rangle ^{ \otimes N}} + \sin \theta {\left| 1 \right\rangle ^{ \otimes N}}$, we have $P_{II}<0$ and
\begin{equation}
P_I(N,n_A)=\sin^2\theta\cos^2\theta n_A!^2(N-n_A)!^2>0 ~~\forall N,n_A\;,
\label{theta-P1}
\end{equation}
which is positive for all partitions $(n_A,n_B)$, verifying the existence of GHZ-type entanglement. Similarly, for the $N$-spin W-state $\left| \psi  \right\rangle  = \frac{1}{{\sqrt N }}\left[ \left| 000 \cdots 01\right\rangle\right. $ $ + \left| {000 \cdots 10} \right\rangle  $ $+  \cdots  $ $+ \left| {001 \cdots 00} \right\rangle  $$+ \left| {010 \cdots 00} \right\rangle $$ \left.+ \left| 100 \cdots 00 \right\rangle  \right]$, we have $P_I<0$ and
\begin{equation}
P_{II}(N,n_A)={\left[ {\frac{{n_A\left( {N - n_A} \right)}}{N}} \right]^2}>0, ~~\forall N,n_A\;.
\label{P2}
\end{equation}

Note that Eq. (\ref{theta-P1}) is maximum for $\theta=\pi/4$, which reflects the fact that the $N$-spin GHZ state is maximally entangled when it is an equal superposition. As $P_I\le 0$ corresponds to separabilty for a particular partition, this observation leads us to the following conjecture for relative measurement of entanglement: {\it The larger positive values of $P_I$ and $P_{II}$  for a fixed set of $(N,n_A)$ indicate more entanglement.} We further notice that

(a) For a given value of $N$ and $n_A$, $P_I$ in Eq. (\ref{theta-P1}) and $P_{II}$ in Eq. (\ref{P2})  set an upper bound, that corresponds to maximally entangled states. For any other states (for example, a mixed state), the values of $P_I$ and $P_{II}$ will be lower than these upper bounds.

(b) $P_I$ and $P_{II}$ are functions of $n_A$, which means that the state is differently entangled for different partitioning. For example, for an $N$-qubit GHZ state, the degree of entanglement is maximum for $n_A=1$ (corresponding to maximum value of $P_I$) and minimum for $n_A=N/2$ (even $N$) or for $n_A=(N-1)/2$ (odd $N$).  On the other hand, for an $N$-qubit W-state, the degree of entanglement is minimum for $n_A=1$ and maximum for $n_A=N/2$ (even $N$) or for $n_A=(N-1)/2$ (odd $N$).

(c) For a given fixed value of $n_A$, $P_I$ monotonically increases as $(N-n_A)!^2$ and $P_{II}$ saturates to $n_A^2$, with increase in $N$. This refers to more GHZ-type entanglement with larger $N$, while degree of W-type entanglement does not change much for large ensemble (i.e., in the limit $N\gg n_A$).

{\it Example 3:} A non-symmetric state  $|\psi\rangle=\left[ {\left| {010} \right\rangle  + \left| {001} \right\rangle } \right]/\sqrt{2}$: This is separable for (1,23) partition as ${\left| {\frac{1}{2}, - \frac{1}{2}} \right\rangle _1}{\left| {1,0} \right\rangle _{2,3}}$ and W-type inseparable for other partitions, viz. (2,13), (3,12). This is complemented by the following results:
\begin{equation}
\begin{array}{l}
{P_I}|_{(12,3)} =  - \frac{1}{2}\;;~~{P_{II}}|_{(12,3)} = \frac{1}{4}\;,\\
{P_I}|_{(1,23)} = 0\;;~~{P_{II}}|_{(1,23)} = 0\;.
\end{array}
\end{equation}

\begin{figure}
\centering
\scalebox{0.3}{\includegraphics{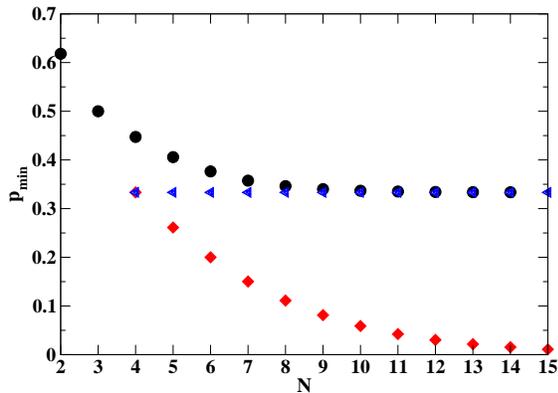}}
\caption{Variation of $p_{\rm min}$ with total number $N$ of spins for $n_A=1$ (black filled circles) and for $n_A,n_B\ne 1$ (red filled squares). The blue filled triangles denote the lower limit for 2-spin Werner state, as obtained by Peres \cite{peres}.}
\end{figure}

{\it Example 4:} Werner state : This is a mixture of a maximally entangled state $|\psi\rangle$ (e.g., an $N$-spin GHZ state) and a fully separable state $\hat{I}$ (the identity operator) as
\begin{equation}
\rho=p|\psi\rangle\langle \psi|+\frac{1-p}{2^N}\hat{I}\;,
\end{equation}
which is entangled for $0<p\le 1$. We find the minimum values of $p$, for which the above state is entangled according to our criterion. We choose to use Eq. (\ref{class1}) and find two different results, as displayed in Fig. 1: (a) For $n_A=1$ and $n_B=N-1$, the $p_{\rm min}$ converges to 1/3. This is similar to the result, obtained by Peres, for 2-spin Werner states. (b) For any other $n_A,n_B\ne 1$, the $p_{\rm min}$  converges to zero for large $N$. Clearly, for a fixed $N$, the value of $p_{\rm min}$ depends upon partitioning. As an example, for $N=4$, the partitioning with $n_A=1$ leads to $p_{\rm min}=0.4472$, while with $n_A=2$, $p_{\rm min}$ becomes 0.3333. This further strengthens our conjecture, as the state is differently entangled for different partitions. We find that for partitioning (b), $p_{\rm min}=[2^{(N-2)/2}+1]^{-1}$.

Though, this is weaker than the original Peres-Horodecki criterion, which shows $p_{\rm min}=[2^{N-1}+1]^{-1}$,
simplicity and uniqueness of our criterion make it favourable in the spin experiments, as it requires measurement of {\it only a few spin-spin correlations\/}. For example, for a 2-spin GHZ-type state [i.e., the Bell state $(|00\rangle+|11\rangle)/\sqrt{2}$], the class 1 inequality (\ref{class1}) reads as $P_I(2,1)=\langle S_+^AS_+^B\rangle\langle S_-^AS_-^B\rangle -\langle S_-^AS_+^AS_+^BS_-^B\rangle$, where
\begin{eqnarray}
\langle S_+^AS_+^B\rangle\langle S_-^AS_-^B\rangle&=&\frac{1}{16}\left[\left(\langle S_x^AS_x^B\rangle-\langle S_y^AS_y^B\rangle\right)^2\right.\nonumber\\
&+&\left.\left(\langle S_x^AS_y^B\rangle+\langle S_y^AS_x^B\rangle\right)^2\right]\;,\nonumber\\
\langle S_-^AS_+^AS_+^BS_-^B\rangle&=& \frac{1}{4}\langle \hat{\bf 1}-S_z^A+S_z^B-S_z^AS_z^B\rangle\;.
\end{eqnarray}
One can derive similar expressions for other combinations of $(N,n_A)$ also.


In conclusion, we presented a new set of criteria, which are sufficient for inseparability of multi-spin systems, based on simple spin-spin correlation measurements and scalable to large $N$.  These are particularly useful to identify rich structure of entanglement between two arbitrary partitions, when single-spin addressing is not possible, e.g., in atomic ensembles. We have illustrated our criteria for different inequivalent classes of $N$-spin symmetric and non-symmetric states as well as states with both odd and even $N$. These also provide a relative quantification of entanglement in two different states, and in a state with different partitions. Since this reveals the any-partition biseparability, it is more relevant in the context of  distillable entanglement \cite{cirac99}. However, detection of bound entanglement is still to be studied using our criterion.

The author thanks Dr. Shubhrangshu Dasgupta for his insightful comments on this work. She also acknowledges financial support under Fast Track Scheme from Department of Science and Technology, Govt. of India (Project No. SR/FTP/PS-71/2009), during this work.

\end{document}